\documentclass[twocolumn,showpacs,preprintnumbers,amsmath,amssymb]{revtex4}

\usepackage{graphicx}
\usepackage{dcolumn}
\usepackage{bm}

\begin{document}


\title{Quasi-continuum approximation to the Nonlinear Schr\"odinger equation with Long-range dispersions} 

\author{Alain~M.~Dikand\'e}
\affiliation{D\'epartement de Physique, Facult\'e des Sciences, Universit\'e de Sherbrooke J1K2R1 Sherbrooke Qu\'ebec, Canada.}  
\email{amdikand@physique.usherb.ca}
\date{\today}

\begin{abstract}
The long-wavelength, weak-dispersion limit of the discrete nonlinear Schr\"odinger equation with long-range 
dispersion is analytically considered. This continuum approximation is carried out irrespective 
of the dispersion range and hence can be assumed exact in the weak dispersion regime. For nonlinear Schr\"odinger 
equations showing finite dispersion extents, the long-range parameter is still a relevant control parameter allowing to 
tune the dispersion from short-range to long-range regimes with respect to the dispersion extent. 
The long-range Kac-Baker potential becomes unappropriate in this context owing to an "edge anomaly" consisting of vanishing 
maximum dispersion frequency and group velocity(and in turn soliton width) in the "Debye" limit. An improved 
Kac-Baker potential is then considered which gives rise to a non-zero maximum frequency, and allows for soliton excitations with 
finite widths in the nonlinear Schr\"odinger system subjected to the long-range but finite-extent dispersion.  
\end{abstract}
\pacs{03.50.-z, 42.65.Tg, 45.20.-d}

\maketitle
\begin{figure}
\includegraphics[width=3in]{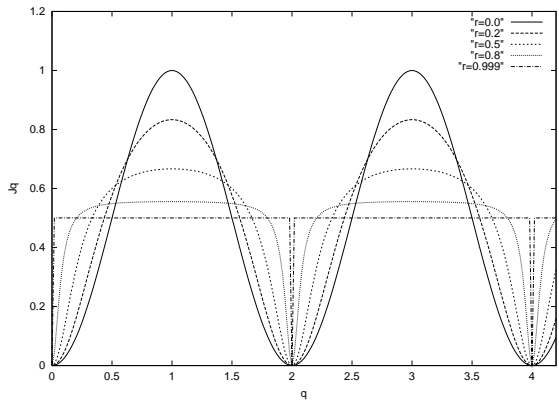}
\caption{\label{fig:dispe1} Plot of the dispersion function versus $q$(in the reduced unit $q/ \pi$) for some values of the long-range parameter $r$. Here the extent of dispersion is taken infinite.}
\end{figure}
\begin{figure}
\includegraphics[width=3in]{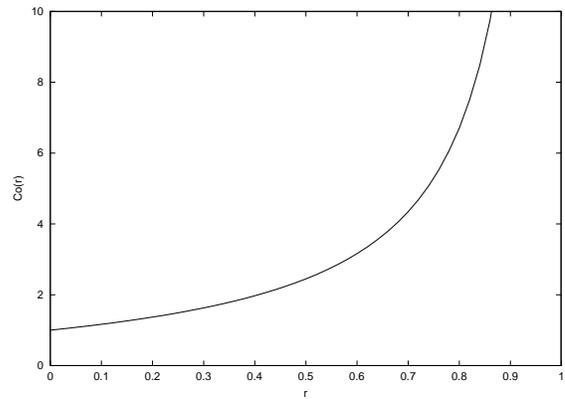}
\caption{\label{fig:vites} Plot of the sound speed versus $r$ for the NLSE with infinite dispersion extent.}
\end{figure}
\begin{figure}
\includegraphics[width=3in]{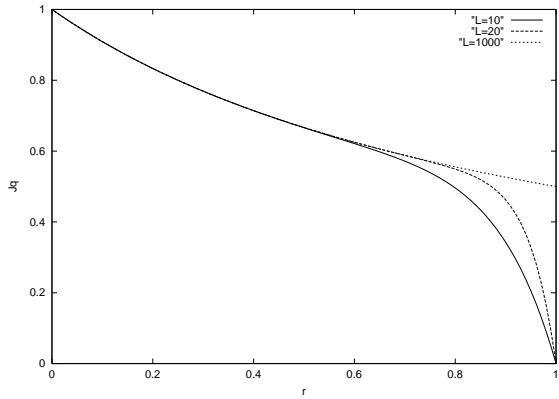}
\caption{\label{fig:dispa1} Dependence of the maximum frequency on the dispersion extent for the LRNLSE with the KB potential. }
\end{figure}
\begin{figure}
\includegraphics[width=3in]{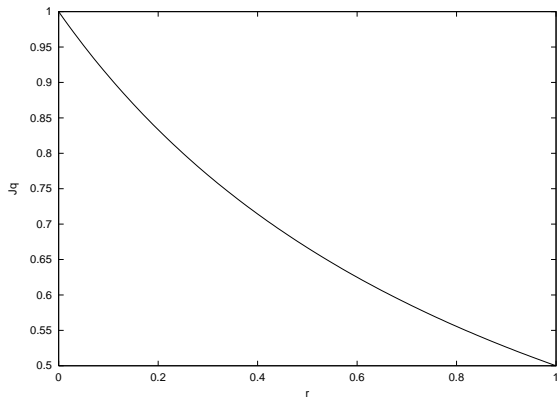}
\caption{\label{fig:dispa2} The maximum frequency of the LRNLSE with the MKB potential. Note the absence of a dependence on the dispersion extent.}
\end{figure}
\begin{figure}
\includegraphics[width=3in]{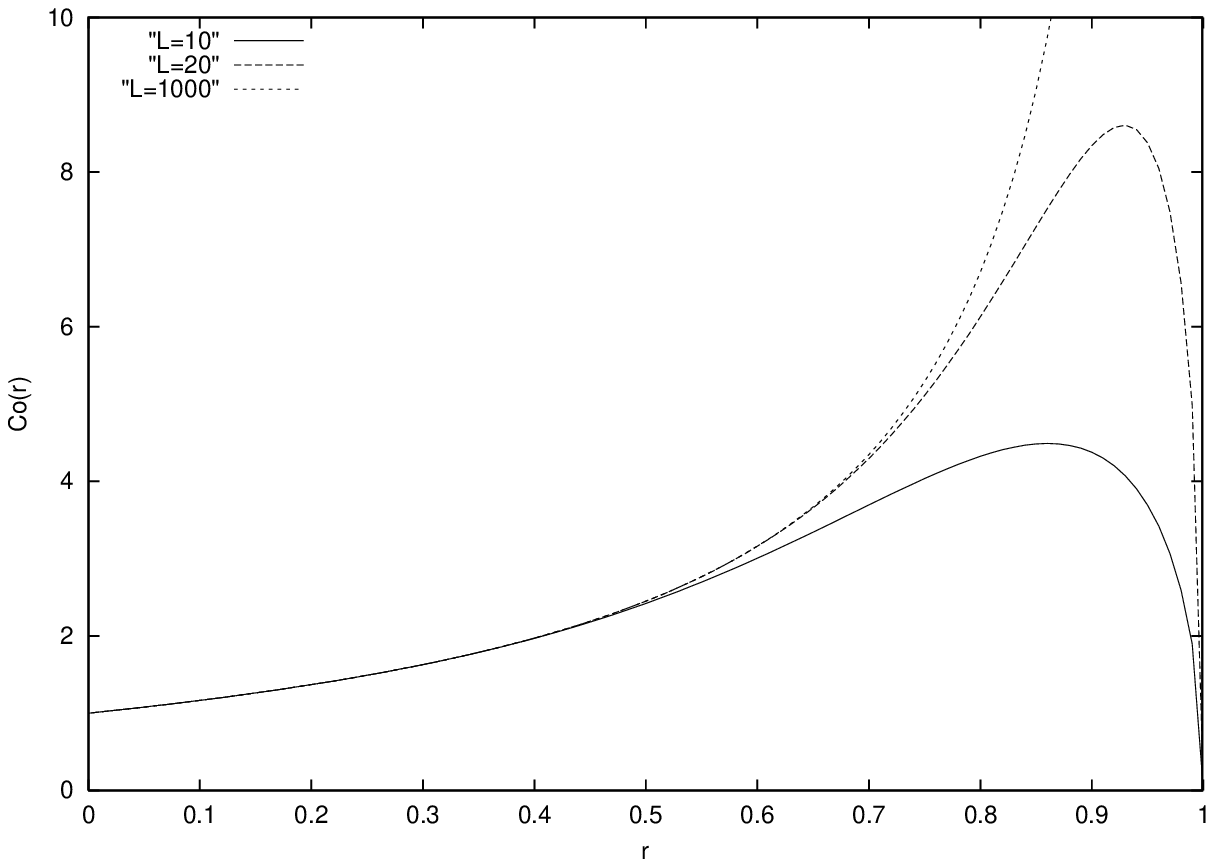}
\caption{\label{fig:vit1} Dependence of the sound speed on the dispersion extent for the KB potential. }
\end{figure}
\begin{figure}
\includegraphics[width=3in]{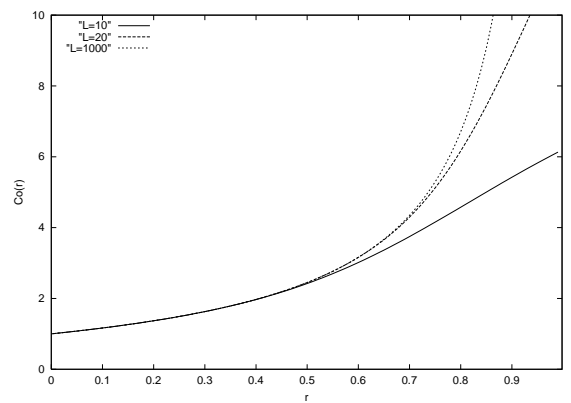}
\caption{\label{fig:vit2} Dependence of the sound speed on the dispersion extent for the MKB potential. }
\end{figure}

Considerable efforts have been devoted to understanding physics of one-dimensional($1D$) systems in 
which dynamical properties 
are dominated by the competition between nonlinearity and dispersion. The Nonlinear Schr\"odinger equation(NLSE) is one 
of most 
investigated nonlinear equations in these contexts, and widespread applications can be found in various fields of condensed matter 
physics. Very recently, a new path was opened in the interest to this equation toward materials with long-range(LR) 
dispersions~\cite{magnus,bishop1}. These LR dispersions are thought to occur in two different kinds namely, intrachain 
dispersions involving long-range interactions among particles of a single discrete chain~\cite{magnus}, and 
interchain dispersions related to the couplings between several $1D$ chains with short-range interactions~\cite{bishop1}. 
Each of these two 
kinds of LR dispersion deserves interest since both lead to two distinct but real physical contexts. Magnetic 
spin chains and Davydov's molecular chains~\cite{davydov,takeno} are two examples of natural systems in which the first 
kind of LR dispersion arises. As for the second, we strongly suspect it may provide an excellent way to model systems made of several 
weakly coupled $1D$ chains. Indeed, if the coupling between the $1D$ chains is not too strong the model can stand as a good 
approximation of a quasi-one-dimensional($Q1D$) material. To this viewpoint, we expect the second kind of LR dispersion to find 
direct applications in nonlinear optics or Bose-Einstein condensate(BEC) systems~\cite{dafolvo,denschlag} where soliton 
compression phenomena~\cite{sipe,koroteev,vlasov} are likely to occur, e.g. resulting from periodic arrangements of dielectric 
elements as in photonic crystals~\cite{yablo,ho,joan} or from the configuration of an optically coupled array of BEC 
lattices. \\
The present work deals with the first kind of LR dispersion. The discrete NLSE in this cas writes~\cite{magnus}:
\begin{equation}
-i\psi_{n t} + \sum_{m\neq n}^L{J_{m-n}(\psi_m - \psi_n)} + g\mid \psi_n \mid^2 \psi_n=0. \label{eqat1}
\end{equation}
where $\psi$ is the complex wavefunction, $g$ is a nonlinear coupling(hereafter assumed positive) 
and $J_{m-n}$ 
is the potential creating the LR dispersion. Equation~(\ref{eqat1}) has been discussed by few 
authors~\cite{magnus} assuming exponential and power-law LR potentials. These works focus on soliton solutions as 
well as on their stability and suggest rich physical properties related to the LR dispersion. In particular, authors of 
ref.~\cite{magnus} combined lattice discreteness and LR dispersion and showed that different soliton regimes may be 
stabilized from the interplay of these two factors. In connection with the variety of stability regimes, the authors 
suggested a switching mechanism by which the transition between different bistable localized states could be monitored by 
tuning parameters of the LR potential. In biophysical systems where inter molecular interactions often display great 
sensitivity to the flexibility of molecular backbones, such a switching mechanism is likely 
to be due to the "contraction-relaxation" features of bond danglings. In this view, provided the bond length is commensurate 
with the wavelength of intra molecular excitations, the discrete NLSE with LR dispersion(LRNLSE) is indeed a good approximation 
for nonlinear molecular vibrations and nonlinear molecular excitons can well be understood as highly discrete and 
localized solutions of this equation with relatively narrow shapes(compared with the lattice constant). However, if the 
bond length is short and the molecular masses large, narrow excitons are less probable since long-wavelength processes 
will tend to dominate. This last situation is the commonly studied one in magnetic spin chains and Davydov models 
in the short-range dispersion regime. The associate NLSE admits single pulse and dark soliton solutions in the continuum 
limit. Strictly speaking, the continuum short-range NLSE(SRNLSE) is the long-wavelength and weak-dispersion limit of the 
discrete SRNLSE~\cite{bishop2}. \\
The goal of the present study is to  point out that an equivalent limit is also accessible for the 
discrete LRNLSE irrespective of the range of lattice dispersion. For this purpose, we will follow a Fourier transform 
method consisting to first 
construct the dispersion function of the LR system, and next extract the continuum NLSE from a second-order expansion 
of the dispersion function. As this expansion does not involves some constraint on the range of lattice dispersion, 
the continuum LRNLSE derived from our long-wavelength and weak-dispersion approximation is indeed an exact equivalent 
of the celebrated continuum SRNLSE. Our interest to the problem was motivated after noting that the dispersion curves 
of the NLSE with exponential LR potentials always displayed the linear shape at small wavectors, and that only their slopes 
were affected by the variation of the LR parameter. In this context, no physical or mathematical constraint prevents us from 
linearizing the dispersion relation of the LRNLSE so far as the long-wavelength limit is concerned. Actually, the variation 
of the slope of the dispersion curve with varying dispersion range is the signature of a varying sound speed. That the entire 
effect of the LR dispersion is confined on the sound speed agrees quite well with the spirit of weak-dispersion 
approximations. Disorder, whcih is another interesting source of residual dispersion, often causes the same effect. We 
would also like to draw attention on similarity between the current problem and that discussed by Konotop~\cite{konotop} 
in terms of the group velocity of a discrete nonlinear monoatomic lattice with Kac-Baker(KB)~\cite{kac,sarker} LR interaction. \\
 For illustrative purpose, we will assume two different LR potentials one of which is the KB potential i.e.($\ell= m - n$):
\begin{equation}
J_{KB}(\ell) = J_o \frac{1-r}{2r} r^{\vert \ell \vert},  \hspace{.1in}  0 \leq r < 1,   \label{kac1}
\end{equation}
 The second LR(MKB) potential is defined as:
\begin{equation}
J_{MKB}(\ell)= J_o \frac{1-r}{2r(1-r^L)}r^{\vert \ell \vert},  \hspace{.1in}  0\leq r \leq 1   \label{kac2}
\end{equation}
In both~(\ref{kac1}) and~(\ref{kac2}), $J_o$ is the potential constant, $r$ is 
the LR parameter and $L$ is half of the spatial extent of the LR dispersion. Indeed, we suppose the actual distance 
on which LR dispersions effectively spread out is not absolutely equal to the total size of the spin or molecular system. For 
instance we can think of an infinite-size system undergoing finite-extent dispersions as it really happens in most cases. 
Neverthess, to avoid losing track of the usual necessary constraints on LR dispersion potentials it is needful to mathematically 
probe the validity of such possibility. For this purpose we sum $J_{m-n}$ over $L$ obtaining: 
\begin{eqnarray}
\sum_{\ell=-L}^L J_{KB}(\ell) &=& Jo(1 - r^L), \nonumber \\
\sum_{\ell=-L}^L J_{MKB}(\ell) &=& Jo.   \label{some}
\end{eqnarray} 
As one sees, the norm of the KB potential is length dependent and becomes constant only in the "thermodynamic" 
limit. The dependence of the total amplitude of the KB potential on $L$, which we term "edge anomaly"~\cite{anomaly}, 
is the main motivation of our interest to the MKB potential. Of course, this edge anomaly does not shades at all the 
rich physics behind the KB model though seeming to restrict its validity to infinite-length materials and systems with 
infinite dispersion extents. Still, making the potential amplitude a function of $L$ as in the MKB potential can turn into a relevant 
advantage for NLSEs with finite dispersion extents. In this case the potential amplitude can be adjusted to a desired 
dispersion extent regardless of whether the total length of the system is finite or not. To this point, note that $J_{MKB}$ reduces 
to $J_o$ in the short-range limit(i.e. $L=1$) irrespective of the value of $r$. On the contrary, for $J_{KB}$ we need to set 
$r=0$. Below a finite-length solution of the 
continuum LRNLSE will be derived and will appear to be a periodic function, the period of which 
can be set equal to $L$. Doing so, we confine solitonic excitations on the effective spatial interval spread by the 
dispersion. Since the LR paramater $r$ allows monitoring the strength of the dispersion regardless of 
its extent, the theory will still be valid whether the system is of finite size or not.  \\
     As we are interested with the nonlinear localized solution of eq.~(\ref{eqat1}), we can write down the wavefunction 
$\psi_n(t)$ as follow:
\begin{equation}
\psi_n(t)= \phi_n \, exp(-i \nu t).     \label{onde}
\end{equation}
Inserting~(\ref{onde}) in~(\ref{eqat1}) we obtain:
\begin{equation}
-\nu \phi_n + \sum_{m\neq n}^L{J_{m-n}(\phi_m - \phi_n)} + g\phi_n^3 =0. \label{eqat2}
\end{equation}    
Since the new wavefunction is only space dependent,~(\ref{eqat2}) is a discrete nonlinear static equation and a spatial 
Fourier transform can be performed. We define the Fourier transform(per unit length) of $\phi_n$ as:
\begin{equation}
\phi_n= \int{dq \, \phi_q \, e^{-i q n d}},  \label{fourier}
\end{equation}
where $d$ is the lattice spacing and $q$ the wavector. Except the LR potential, all coefficients in eq.~(\ref{eqat2}) are 
constant so their Fourier transforms are single inegrals. Remark that the 
nonlinear term of this equation describes an isotropic "three-mode" coupling and as consequence the associate Fourier tranform will be a 
simple integration over the single wavector common to the three coupled fields. The most interesting part of the discrete 
nonlinear static equation is thus the second term whose Fourier tranform writes:
\begin{eqnarray} 
\sum_{m\neq n}^{\infty}J_{m-n}(\phi_m - \phi_n) & =& - \int{dq \, J_q(r) \, \phi_q \, e^{-i q n d}},   \nonumber \\
J_q(q) &=& J(r) \sum_{\ell=1}^L r^{\ell} \left[1 - \cos(q \ell d) \right],   \nonumber \\
                                        \label{dispero}
\end{eqnarray}  
where $J(r)$ can assume two distinct forms in connection with~(\ref{kac1}) and~(\ref{kac2}) i.e.:
\begin{eqnarray}
J_{KB}(r) &=& J_o \frac{1-r}{r},   \nonumber \\
J_{MKB}(r)&=& J_o \frac{1-r}{r(1-r^L)}.           \label{consts}
\end{eqnarray}
The quantity $J_q(r)$ is the function governing the spatial dispersion of the LRNLSE. 
The sum appearing in this function is restricted to $L$. However, it can be 
extended to infinity provided the chain length also is infinite. Since this last context is more familiar, 
let us first examine the case $L \rightarrow \infty$. On fig.~\ref{fig:dispe1} we plot $J_q$ versus $q$ for some 
values of $r$. The just mentioned sum is evaluated numerically for the two LR potentials. 
 According to~(\ref{kac2}), both potentials reduce to the same expression in the infinite-dispersion limit. \\
Fig.\ref{fig:dispe1} shows that $J_q(r)$ remains a sinusoidal function for the whole interval of values of $r$, and  
only slopes of the dispersion curves vary. To check this assertion, a numerical expansion of the dispersion function $J_q$
was carried out up to the second-order term for weak values of $q\, d$. The leading and first-order terms vanish but not the 
second-order term. The coefficient of this last term is nothing else but the square of the sound speed $C_o(r)$~\cite{konotop}, which is 
plotted in fig.~\ref{fig:vites} as a function of the LR parameter $r$. \\
An exact analytical expression of $J_q(r)$ for finite $L$ can be obtained using the following identity:
 \begin{eqnarray}
 V_r(q)&=& \sum_{\ell=1}^L r^{\ell} \cos(q d \ell), \nonumber \\
      &=& \frac{r \cos(\tilde{q}) - r^2 - r^{(L+1)} [\cos((L+1)\tilde{q}) - r \cos(L\tilde{q}) ]}{1 - 2r \cos(\tilde{q}) + r^2}. \nonumber \\
                                                \label{disper1}
\end{eqnarray}
and is given by:
\begin{equation}
J_r(q) = J(r)[V_r(0) - V_r(q)],     \label{eqdisp}
\end{equation}
 We can check that this analytical expression is fully consistent with numerical curves in fig.~\ref{fig:dispe1}. To 
see the effects of finite $L$ on the dispersion of our system, in figs.~\ref{fig:dispa1} 
and~\ref{fig:dispa2} we have drawn~(\ref{eqdisp}) versus $r$ placing ourselves at the edge of the first Brillouin zone i.e. 
$q=\pi$, and using $L=10$, $20$, $1000$. We see that the maximum frequency is sensible to the extent of dispersion for the 
KB potential and not for the MKB potential. However, the two sound speeds(figs.~\ref{fig:vit1} and~\ref{fig:vit2}) fill 
effects of finite values of $L$ but tend to the same limit when $L$ becomes sufficiently large(i.e. infinite). In the case 
of KB potential, there occur maxima in $C_L(r)$ at $r<1$ and the sound speed vanishes as $r\rightarrow 1$ for finite $L$. 
As opposed to this first model, $C_L(r)$ is monotonous in the MKB case and tends to a finite value as $r\rightarrow 1$ when 
$L$ assumes finite values. \\  
 The analytical expression of the sound speed plotted in figs~\ref{fig:vit1} and~\ref{fig:vit2} was derived 
from~(\ref{eqdisp}) by an analytical expansion of $J_q(r)$ to the second order for weak values of the product $q \, d$. Since here 
too the leading and first-order terms vanish, we are only left with the second-order term proportional to $q^2$. Strictly, this 
expansion corresponds to a continuum-limit approximation for long-wavelength($n d \rightarrow x$) and 
weak-dispersion ($q d << 1$) excitations. Thus, with regard to the 
minus sign in front of the right-hand side of the first equation in~(\ref{dispero}), an inverse Fourier tranform of the 
$q^2$ term produces the following continuum nonlinear second-order equation:
\begin{equation}
C_L^2(r) \phi_{xx} -\nu \phi_n + g \phi_n^3 =0. \label{eqat3}
\end{equation}    
The square of the sound speed $C_L^2(r)$ is proportional to the second-order derivative of $J_q(r)$ i.e.: 
\begin{equation}
C_L^2(r) = (1/2) \partial_{qq} \Omega^2(q) \vert_{q \rightarrow 0}. \label{vite}
\end{equation} 
It is useful to note that the discrete LRNLSE~(\ref{eqat1}) itself could be treated in a similar manner and would lead to the 
usual form of continuum NLSE. \\
For positive values of $g$ and $\nu$, equation~(\ref{eqat3}) admits finite-length soliton solutions~\cite{dika} given in 
terms of Jacobi Elliptic functions~\cite{elliptic}:
\begin{eqnarray}
\phi(x) &=& \phi_o cn\left(\frac{x}{\ell_o}\vert \kappa \right), \hspace{.2in} \kappa \geq 1, \nonumber \\
\phi_o &=& \sqrt{\frac{2 \kappa^2}{2\kappa^2 - 1}} \left(\frac{\nu}{g}\right)^{1/2}, \hspace{.2in} \ell_o^2 = \frac{C_L^2(r)}{\nu}(2\kappa^2-1). \nonumber \\
                 \label{snoid}
\end{eqnarray}
$cn$, which is the Jacobi Elliptic function, is periodic in $x$ with a period $L_o= 4\ell_o K$ where $K$ is the complete Elliptic integral of 
the first kind. In general, cnoidal waves provide adequate representations of the soliton solutions of finite-length systems 
in connection with the finite magnitude of $L_o$. In this viewpoint, we can decide to set $L=L_o$ which amounts to 
confine all solitonic excitations within the interval spread by the LR dispersion. In the limit $\kappa=1$,~(\ref{snoid}) 
turns to the single-pulse soliton:
\begin{eqnarray}
\phi(x) &=& \phi_o sech\left(\frac{x}{\ell_o} \right)\nonumber \\
\phi_o &=& \sqrt{\frac{2\nu}{g}}, \hspace{.2in} \ell_o^2= \frac{C_L^2(r)}{\nu}  \label{pulse}
\end{eqnarray}
and $L_o \rightarrow \infty$. However, $L$ can be kept finite in this limit. According to the dependence of 
$\ell_o$ on the sound speed, we expect the pulse width to assume behaviours of $C_L(r)$ with respect to the LR parameter $r$. 
On fig.~\ref{fig:vites}, $C_{\infty}(r)$ was an infinitely increasing function of $r$. This means the pulse width is 
quickly diverging as $r\rightarrow 1$ for infinite dispersion extents. Following the effect of an assumption of finite 
values of $L$ on the sound speed, finite-width single-pulse solitons are likely to be observed in a NLSE with LR dispersion and 
corresponds, in a physical context, to a system in which the extent of dispersion does not follow the system size. 
However, since the sound speed always goes to zero as $r\rightarrow 1$ for the KB potential, this model is relatively less 
suitable in such physical contexts as opposed to the apparently advantageous features of the MKB potential.
   
\begin{acknowledgments}
I would like to express thanks to Dr. T. C. Kofan\'e from the University of Yaound\'e, cameroon for numerous advices. Part of this work 
was carried out as the author was guest at the Abdus Salam International Centre for Theoretical Physics Trieste, Italy.
\end{acknowledgments}


\begin{thebibliography}{}
\bibitem{anomaly} For finite values of $L$, the total amplitude of the Kac-Baker potential is zero in the limit $r\rightarrow 1$. 
Consequent upon this behaviour are zero maximum frequency and sound speed hence "edge anomalies".
\end{thebibliography}
\end{document}